\documentclass{sigchi}

\RequirePackage{snapshot}

\toappear{
Permission to make digital or hard copies of all or part of this work for personal or classroom use is granted without fee provided that copies are not made or distributed for profit or commercial advantage and that copies bear this notice and the full citation on the first page. Copyrights for components of this work owned by others than the author(s) must be honored. Abstracting with credit is permitted. To copy otherwise, or republish, to post on servers or to redistribute to lists, requires prior specific permission and/or a fee. Request permissions from Permissions@acm.org.
}

\usepackage{graphics} 
\usepackage[T1]{fontenc}
\usepackage{txfonts}
\usepackage{mathptmx}
\usepackage{color}
\usepackage[usenames,dvipsnames,table]{xcolor}
\usepackage{booktabs}
\usepackage{textcomp}
\usepackage{ccicons}  
\usepackage[utf8]{inputenc} 

\usepackage{morefloats}
\usepackage{gensymb}
\usepackage{subcaption}
\usepackage{amsmath}
\usepackage{amsfonts}
\usepackage{amssymb}
\usepackage{hyperref}
\usepackage[all]{hypcap}  

\newcommand{\figref}[1]{Fig.~\ref{fig:#1}}

\def\plaintitle{To Walk or Not to Walk: Crowdsourced Assessment of External Vehicle-to-Pedestrian Displays}

\def\emptyauthor{} \def\plainkeywords{Vehicle-to-pedestrian communication, external HMI, autonomous vehicles,
  crowdsourced design, shared road users, vulnerable road users, pedestrian communication, communication displays.}

\makeatletter
\def\url@leostyle{%
  \@ifundefined{selectfont}{
    \def\UrlFont{\sf}
  }{
    \def\UrlFont{\small\bf\ttfamily}
  }}
\makeatother
\def\pprw{8.5in}
\def\pprh{11in}

\definecolor{linkColor}{RGB}{6,125,233}
\hypersetup{%
  pdftitle={\plaintitle},
  pdfauthor={\emptyauthor},
  pdfkeywords={\plainkeywords},
  bookmarksnumbered,
  pdfstartview={FitH},
  colorlinks,
  citecolor=black,
  filecolor=black,
  linkcolor=black,
  urlcolor=linkColor,
  breaklinks=true,
}

\begin{document}

\title{To Walk or Not to Walk: Crowdsourced Assessment of \vspace{0.1in}\\ External Vehicle-to-Pedestrian Displays\vspace{0.1in}}

\newcommand{\authorspace}{\hspace{0.1in}}
\author{
Lex Fridman \authorspace
Bruce Mehler \authorspace
Lei Xia \authorspace
Yangyang Yang \authorspace
Laura Yvonne Facusse \authorspace
Bryan Reimer\vspace{0.05in}\\
\affaddr{Massachusetts Institute of Technology}\\
}

\maketitle

\begin{abstract}
  Researchers, technology reviewers, and governmental agencies have expressed concern that automation may necessitate
  the introduction of added displays to indicate vehicle intent in vehicle-to-pedestrian interactions. An automated
  online methodology for obtaining communication intent perceptions for 30 external vehicle-to-pedestrian display
  concepts was implemented and tested using Amazon Mechanic Turk. Data from 200 qualified participants was quickly
  obtained and processed. In addition to producing a useful early-stage evaluation of these specific design concepts,
  the test demonstrated that the methodology is scalable so that a large number of design elements or minor variations
  can be assessed through a series of runs even on much larger samples in a matter of hours. Using this approach,
  designers should be able to refine concepts both more quickly and in more depth than available development resources
  typically allow. Some concerns and questions about common assumptions related to the implementation of
  vehicle-to-pedestrian displays are posed.
\end{abstract}


\keywords{\plainkeywords}

\section{Introduction}\label{sec:introduction}

The introduction of semi-automated and automated driving technologies into the vehicle fleet is often seen as having the
potential to decrease the overall frequency and severity of crashes and bodily injury \cite{national2017federal}. At the
same time, there is some concern that the transition from manually controlled to technology controlled vehicles could
have unintended consequences. One such concern is in the area of communication of intent between automated vehicles and
shared road users, particularly pedestrians
\cite{keferbock2015strategies,lagstrom2015avip,lundgren2017will,matthews2015intent,mirnig2017three}. One perspective
asserts that human driven vehicle-pedestrian communication often involves hand and body gestures, as well as eye contact
(or avoidance of), when vehicles and pedestrians come together in interactions such as those occurring at crosswalks
where miscommunication can easily elevate risk. The question is then posed as to what will replace these forms of
communication when a human is no longer actively driving the vehicle?

One approach might be to explore technologies such as the Wi-Fi application proposed in \cite{anaya2014vehicle} or other
Vehicle-to-Entity (V2X) communications that alert pedestrians of potential conflict situations. However, most proposed
solutions focus on external vehicle displays as replacements for human-to-human visual engagement. Google drew attention
to this approach by filing a patent for messaging displays for a self-driving vehicle that included the concept of
electronic screens mounted on the outside of the vehicle using images such as a stop sign or text saying ``SAFE TO
CROSS'' \cite{urmson2015pedestrian}. Drive.ai, a self-driving technology start-up, released an illustration of a
roof-mounted display screen concept that combined an image of a pedestrian on a cross walk and the words ``Safe to
Cross'' \cite{knight2016selfdriving}. Matthews and Chowdhary \cite{matthews2015intent} describe an LED display for an
autonomous vehicle that might display messages such as ``STOP'' or ``PLEASE CROSS'' when a pedestrian is
encountered. Mirnig and colleagues \cite{mirnig2017three} briefly describe several visual display strategies that they
describe as being informed from human-robot interaction principles. Automotive manufacturers have also proposed design
visions such as the Mercedes F015 concept car using lighted displays on the front grill and a laser projection of an
image of a crosswalk on the roadway in front of car \cite{keferbock2015strategies}. A Swedish engineering company has
proposed a lighted grill design that ``smiles'' at pedestrians to indicate they have been detected and it is safe to
cross in front of the vehicle \cite{peters2016selfdriving}.

\subsection{Early Stage Design Assessment}

Careful and extensive testing vehicle-to-pedestrian communication concepts under real-world conditions and with a broad
demographic sampling would seem to be indicated before a design is put into general use due the potentially safety
critical implications of miscommunication. Given the inherent costs of real-world validation testing, efficient methods
for early stage concept assessment are highly desirable for narrowing in on designs that are promising and setting aside
those less likely to prove out. Further, early stage methods that make it practical to test a large number of minor
design variations increase the probability of elucidating subtle considerations that may lead to optimized
implementations.

Wizard of Oz approaches to assessing how pedestrians might interact with automated vehicles and various external design
concepts have been reported \cite{habibovic2016evaluating,lagstrom2015avip} and \cite{doric2016novel} have described a
virtual reality based pedestrian simulator. While these methods are less intensive than full scale field testing, they
still require significant effort and strategic choices need to be made in selecting concepts to test at this level.

To gather data on the extent to which pedestrians might be uncomfortable and uncertain about whether they should cross
in front of a vehicle if they were unable to make eye contact with the driver (which was presumed to be more likely in
vehicles under autonomous control), Lagstr{\"o}m and Lundgren \cite{lagstrom2015avip} presented participants with a set
of five photographs ranging from an image of the person in the driver's seat holding onto the steering wheel and looking
into the camera to one where the individual in the ``driver's'' seat appeared to be asleep. Participants were asked to
image that:

You are walking through a city center and are just about to cross an unsignalized zebra crossing. A car has just stopped
and you look into the car before passing the crossing, you see what is shown on the picture. How do you feel about
crossing the road?

Participants then were asked for each to respond for each image whether they would cross immediately and made ratings of
their likely emotional reactions. Lagström and Lundgren interpreted the responses as supporting a concern that there may
be a risk of misinterpretation on the part of pedestrians that observing a ``driver'' occupied by activities such as
reading or sleeping as indicating that the vehicle was not about to move. They note that this might be a wrong
interpretation for an automated vehicle and thus some indication of whether a vehicle is in autonomous mode and what its
intentions may be desirable.

The present study employed a data gathering approach that is conceptually similar to Lagström and Lundgren's in that it
presented participants with multiple pictures of a vehicle and asked if it was safe to cross in front of the
vehicle. However, there were two key differences. First, the goal of the assessment was to evaluate design concepts for
communication from the vehicle to a pedestrian whether the pedestrian should cross or not. Second, an automated online
presentation methodology was employed that supported efficient presentation of a large number of images across a larger
sample of participants.

\section{Methods}

Amazon Mechanical Turk (MTurk) was used for data collection. MTurk is an internet based, integrated task presentation
and participant compensation system. It provides access to a large potential participant pool at modest cost per
participant. It has been reported that MTurk has good performance, especially on psychology and other social sciences
research, since participants are diverse and more representative of a non-college population than traditional samples
\cite{buhrmester2011amazon,paolacci2010running}. Participants who take frequently part in MTurk tasks are commonly
referred to as ``Turkers''.

One of the challenges of constructing a statistically meaningful MTurk experiment is the ability to filter out any
responses by Turkers that were not made with their full attention on the task and representative of a ``best
effort''. We used two types of filtering: (1) accepted only select Turkers with a proven track record on MTurk (see
Participants subsection) and (2) the insertion of ``catch'' stimuli for which there is a ``correct'' answer (see Stimuli
subsection).

A second challenge of setting up a successful MTurk experiment is making to scalable to hundreds or thousands of
participants. To this end we implemented a Python framework that created, configured, and served the stimuli in
randomized order on a HTML front-end. An asynchronous Javascript (Ajax) communication channel stored the responses in a
PostgreSQL database through a PHP-managed backend. This framework allowed for robust, concurrent collection of the
dataset underlying this work in just a few hours. Moreover, it allowed for efficient validation of the result and
possible future scaling of the number of Turkers and stimuli.

\subsection{Participants}

To take part, participants had to be experienced Turkers with a minimum of 1,000 previous HITs (a measure of previous
experience where a HIT represents a single, self-contained task that an individual can work on, submit an answer, and
collect a reward for completing) and a 98\% or higher positive review rating. Data collection continued until 200
Turkers completed the full experiment by providing a response to each of the 30 stimuli. According to tracking of IP
addresses, the majority of participants were from the USA and India, and approximately matched the distribution reported
by \cite{ross2009turkers} where 57\% of Turkers were from USA and 32\% were from India. Compensation was at approximately
\$15/hour based on a conservative estimate of a pace necessary to complete the full experiment. This rate is above the
compensation of \$2-3/hour commonly provided on MTurk.

\subsection{Stimuli / communication design elements}

\begin{figure}[ht!]
  \centering
  \includegraphics[width=\columnwidth]{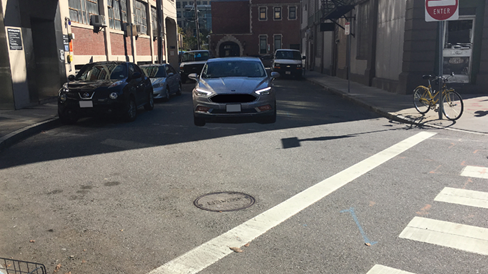}
  \caption{The driver was not visible in the base image used to create the designs due to lighting and reflection angles.}
  \label{fig:base}
\end{figure}

The base photograph used to create the stimuli (see \figref{base}) was of a late model passenger sedan on a one-way urban street
approaching an uncontrolled intersection/crosswalk (no traffic light). Under the lighting conditions the driver is not
visible.

\begin{figure}[ht!]
  \centering
  \includegraphics[width=\columnwidth]{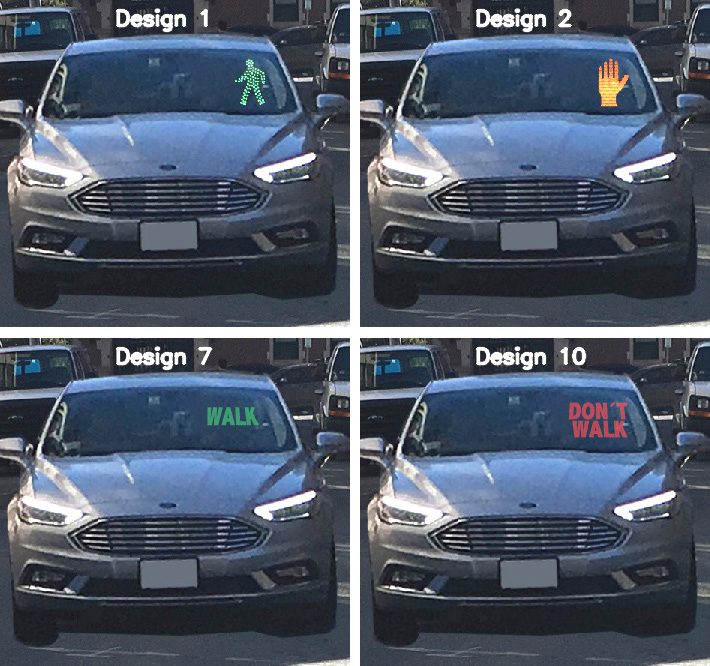}
  \caption{Four of the designs tested (out of 30 total in \figref{designs-all}) shown in cropped, close-up views. Presentation to
    participants used the entire image shown in \figref{base} at a 1280 pixels wide and 720 pixels tall
    resolution. These 4 designs performed significantly better than the other 26 at communicating their intent as shown
    in \figref{plot-all}.}
  \label{fig:four-designs}
\end{figure}

The stimuli were created by superimposing each of the 30 designs onto the base image. Every design had an animated
element in that it was either flashing or playing through a sequence of animation frames. \figref{four-designs} shows illustrative
snapshots of four of the designs. A video of all 30 final stimuli is provided as supplementary material. The size of the
stimuli presented to each Turker was 1280 pixels wide and 720 pixels tall. Turkers with screen resolutions below this
size were automatically detected and could not participate in the experiment.

``Catch'' stimuli were created that, instead of a design, showed instructions on what to respond (e.g., Yes, No). Only
responses provided by Turkers who passed these catch stimuli were included in the resulting dataset. Given the filtering
in the Turker selection, 100\% of the Turkers who completed the entire experiment responded to the catch stimuli
correctly.

\begin{figure*}[htp!]
  \centering
  \includegraphics[width=\textwidth]{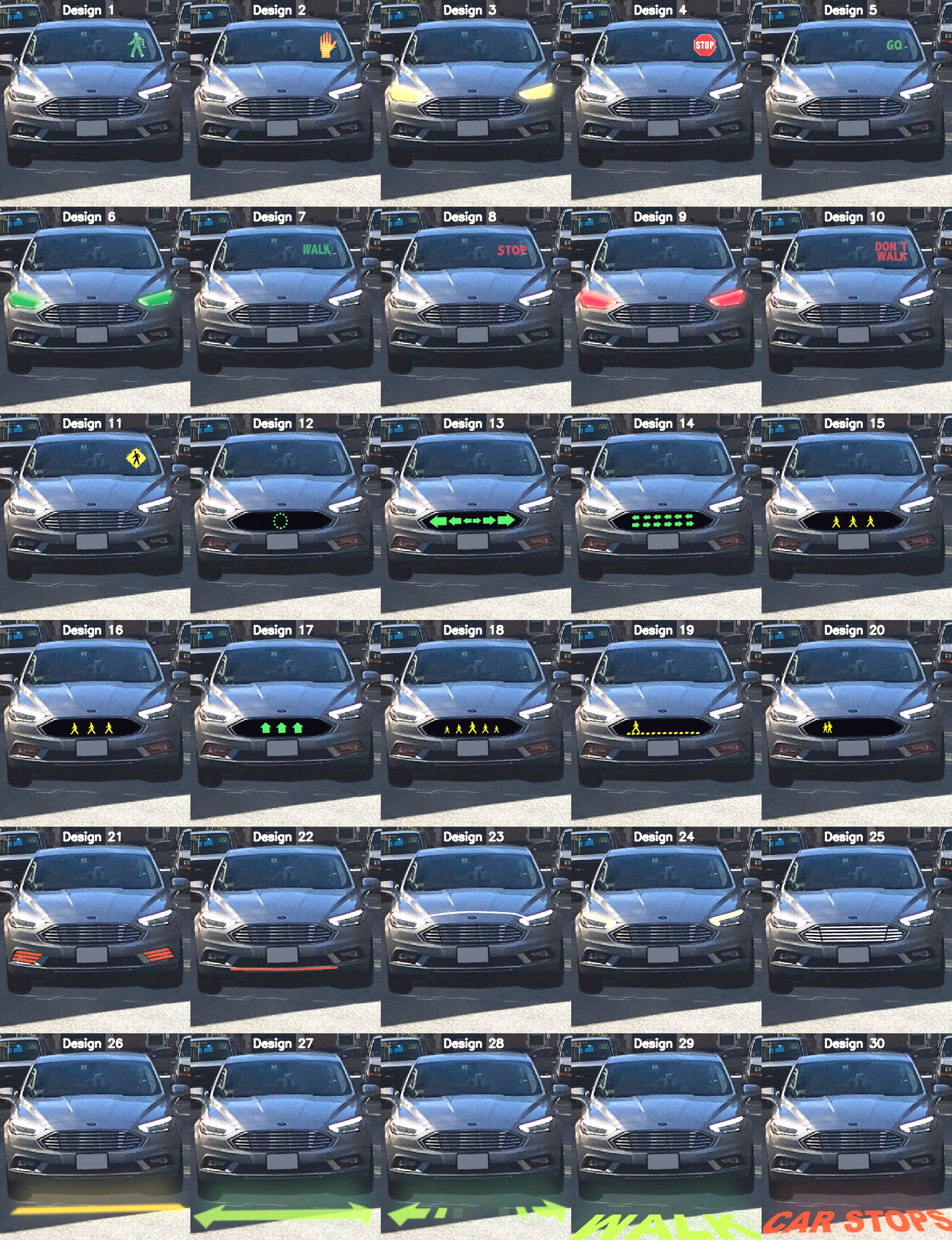}
  \caption{Frames from each of the 30 animated designs presented to participants, shown here in cropped, close-up views.}
  \label{fig:designs-all}
\end{figure*}

\subsection{Procedure}

Participants were presented with introductory text informing them that they would be presented with a series of images
of a vehicle approaching a cross walk. They were to imagine that they were a pedestrian viewing the approaching vehicle
and decide if it was safe to cross. Response options were: Yes, No, and Not Sure. The presentation order for the images
was randomly shuffled for each participant to control for order effects.

Each of the stimuli was animated on screen indefinitely until the Turker provided a response. Response timing
information was recorded, but analysis did not reveal any meaningful patterns or correlation between designs and
response dynamics.

\begin{figure*}[htp!]
  \centering
  \includegraphics[width=\textwidth]{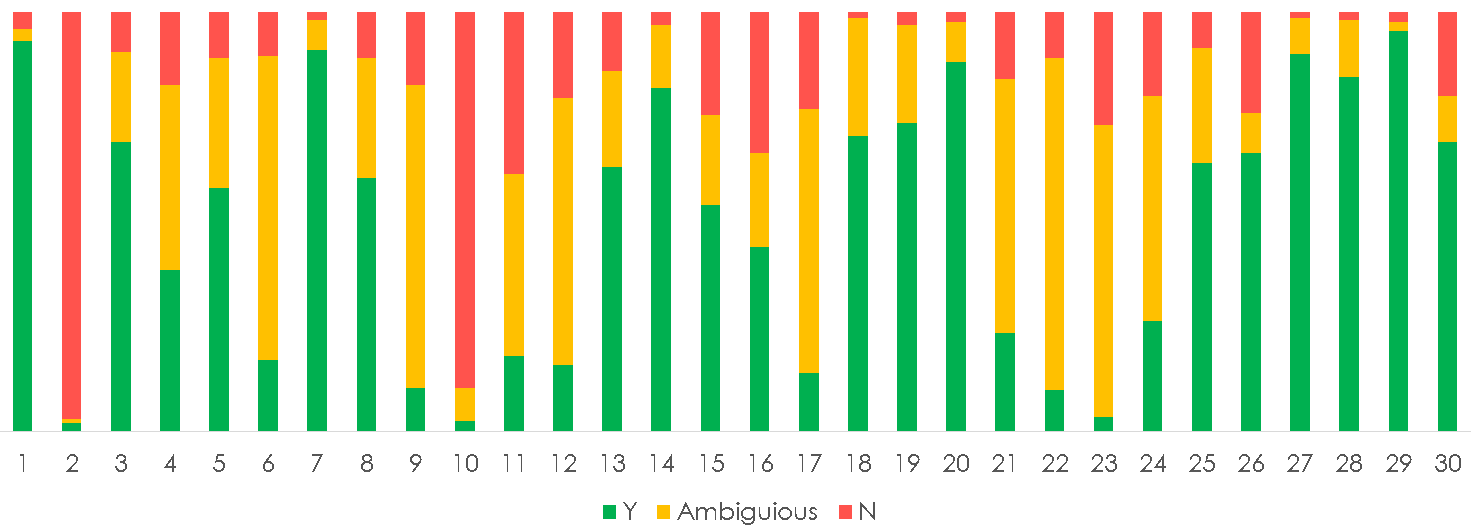}
  \caption{Proportion of 200 participants interpreting each message as indicating it was safe to walk (Yes), yellow
      indicating not sure (Ambiguous), and red interpreting the message as not to walk (No). Designs 2 and 10 were
      intended to communicate not to walk (No).}
  \label{fig:plot-all}
\end{figure*}

\section{Results}

A few of the concepts showed a high degree of match between the designers' intent and participants' interpretation. For
the examples shown in Figure 2, designs 1 and 7 received a high percentage of responses that it was safe to walk;
designs 2 and 10 received a high percentage of responses that it was not safe to walk. Participants' interpretation of
the communication intent on the part of the vehicle for each of the 30 designs is shown in Figure 3; the proportion of
the sample that rated each design as indicating it was safe to walk is colored coded in green, were unsure about the
intent in yellow, and interpreted the message as indicating they should not walk is shown in red. It can be observed in
Figure 3 that the two designs intended to communicate that pedestrians should not walk (2 and 10; images in Figure 2)
match relatively well with participants' interpretations. The degree of successful communication of designer intent was
much more varied for the designs intended to indicate that it was safe to walk; the interpretation ratings of these
designs are broken out in Figure 4. The responses for concepts the designers thought would be ambiguous are shown in
Figure 5.

\section{Discussion}

The external vehicle display concepts intended to communicate vehicle intent to pedestrians were developed by a team of
graduate design students (Xia, Yang, and Facusse) as part of a course project. As external advisors and collaborators,
the remaining authors attempted to guide the students understanding of the potential need for external vehicle displays,
while minimize the amount of input on the designs. The advisors, focused on the design of the web based assessment
methodology, technical aspects of the MTurk implementation, and empirical data collection to provide the students with a
data set that could be used to evaluate assumptions about design elements. The ability to ``risk'' resources in testing
such a large number of design concepts was only practical through the use of a relatively low cost and low time
intensive prototyping methodology such as was explored here. A strong focus of this work was allowing the graduate
design students flexibility to consider the advantages and disadvantages of various design approach and gain rapid
consumer facing feedback on the appropriateness of their decisions.

Twenty of the designs (1, 7, 11-24, 26-30) were created with the intent of communicating that it is safe to walk without
explicitly intended ambiguity. Of these, six of the designs obtained an 80\% or greater match and none of the designs
showed universal agreement. For 8 of the 20 ``walk'' designs, more than half the participants found the message unclear
or misinterpreted the message as ``don't walk''. The 2 ``don't walk'' designs faired generally better, although the
interpretive match was still not universal – which is of particular concern for a safety critical communication in which
nearly, of not 100\%, correct interpretation is needed. Clarity and unambiguity will be critical if external
communication displays are to achieve the goal of building psychological trust between human and machine
\cite{keferbock2015strategies}.

The presence of uncertainty and misinterpretation with all of the designs tested suggests some potential concern around
the concept of ``needing'' to employ external communication signals in automated vehicles intended for public roadways
beyond those already used in non-automated vehicles (e.g., turn signals, brake lights, and vehicle kinematic
cues). Lagström and Lundgren (\cite{paolacci2010running}; see also \cite{lundgren2017will} present a substantive series
of small studies that document pedestrians' desire to understand a driver / vehicle's intent, and explore a creative
design concept involving a row ``movable'' light bar elements at the top of the front windshield to communicate several
messages (e.g. ``I'm about to yield.'', ``I'm about to start.''). After training in the intended meaning of the
messages, all 9 participants in the final test phase were able to correctly report the intention of all of the messages
except for the message intended to indicate that the vehicle was in automated mode. What is not clear is how an
untrained population would interpret the messaging and whether the net result over time would be greater comfort with
automated vehicles and an overall safety benefit for pedestrians.

In another very detailed study, Clamann and colleagues \cite{clamann2017evaluation} tested a variety of designs
including a mock automated van with a prominently mounted, large LCD display employing what would appear to be
relatively apparent walk/don't walk graphics (walking figure with and without a diagonal line across the image). It was
concluded that while a large number of participants felt that additional displays will be needed on automated vehicles,
most appeared to ignore the displays and rely on legacy behaviors such as gap estimation and inferring vehicles'
approaching speed (collectively kinematics) in making decisions on whether or not to cross the road. In an interview
\cite{lefrance2016pedestrians}, Clamann observed that the displays tested were ``as effective as the current status quo
of having no display at all.''

The senior researchers on this paper have, as part of a different project, been involved in extensive observation of
pedestrian-vehicle interactions \cite{toyoda2017understanding}. During these observations, we have increasingly
developed the impression that pedestrians may take their primary communication cues from overt vehicle kinematics more
often than actually depending on eye-contact or body gestures to make judgements about vehicle intent and that multiple
attributes may be used to predict intent. As such, vehicle systems may be developed to be responsive to pedestrian
movements. Thus, we see it as still an open research question as to whether new external displays are necessarily a
priori answer to improving communication of intent. We are in full agreement with Clamann \cite{lefrance2016pedestrians}
that careful, detailed evaluations need to be carried out to make sure that displays and signals work as intended before
they are standardized, mandated or released in any production fashion.

It is clear, that unanticipated consequences can easily occur if a pedestrian in the dilemma zone (stepping off the curb
into the flow of traffic) pauses for even a moment to perceive, read or interoperate the intent of external
communication devices. As such, while benefits of external vehicle displays could easily improve the communication of
intent in a ``trained'' or ``habituated'' population, without nearly ubiquitous understanding risks could easily
increase. Furthermore, a transition period during which a mixed population of vehicles with and without communication
devices, and a mixed set of educated and non-educated pedestrians could be detrimental to short term safety making the
societal hurdles to successful adoption of a new technology more difficult.

\section{Conclusion}\label{sec:conclusion}

Experience implementing the assessment methodology described here in MTurk demonstrates that this approach can be
applied in a cost effective manner for identify design concepts that may be appropriate for more detailed development
and testing. Since a relatively large number of elements or minor variations can be tested through a series of MTurk
runs in a matter of hours (as opposed to weeks or months for focus groups or experimental simulation or field testing),
designers should be able to refine concepts both more quickly and in more depth than available development resources
typically allow. Factors that are often difficult to explore during design phases (e.g. culture, demographic, prior
mental model, etc.) can be factored in early in the process. It is worth noting that there is nothing in this method of
early stage design development that is limited to the messaging application explored in this study; it should be equally
applicable to work on other design elements such as interior interface icons, graphics, gages, and other forms of
information presentation in automotive, consumer electronics, advertising and other domains.

\section{Limitations}

As noted, the majority of participants were from the USA and India, so it is unknown the extent to which the findings
for specific design elements generalize to other regions. A single vehicle type and setting were assessed. The Turker
sample was motivated to pay attention to details of the images and presumably not distracted (e.g. talking on a phone,
etc.), thus correct detection of communication intent may have been greater than might be obtained under real-world
conditions and may represent something approximating best case evaluations. With these considerations in mind, the
methodology explored here is likely to be most useful for rapidly identifying designs or design elements that are
promising for further investigation as opposed to use for late stage validation.

\section{Acknowledgment}

Authors Xia, Yang, and Facusse would like to thank their course instructor, Professor Leia A Stirling, for the guidance
on human centered design and research methodology. Support for this work was provided by the Toyota Class Action
Settlement Safety Research and Education Program. The views and conclusions being expressed are those of the authors,
and have not been sponsored, approved, or endorsed by Toyota or plaintiffs' class counsel.

\bibliographystyle{sigchi}
\bibliography{walk}

\end{document}